# A Graphical Password Based System for Small Mobile Devices

**Wazir Zada Khan[1], Mohammed Y Aalsalem[2] and Yang Xiang[3]**

[1] School of Computer Science, University of Jazan
Jazan, PoBox # 114, Kingdom of Saudi Arabia

[2] School of Computer Science, University of Jazan
Jazan, PoBox # 114, Kingdom of Saudi Arabia

[3] School of Information Technology, Deakin University, Australia
221 Burwood Highway, Burwood, VIC 3125, Australia

**Abstract**

Passwords provide security mechanism for authentication and protection services against unwanted access to resources. A graphical based password is one promising alternatives of textual passwords. According to human psychology, humans are able to remember pictures easily. In this paper, we have proposed a new hybrid graphical password based system, which is a combination of recognition and recall based techniques that offers many advantages over the existing systems and may be more convenient for the user. Our scheme is resistant to shoulder surfing attack and many other attacks on graphical passwords. This scheme is proposed for smart mobile devices (like smart phones i.e. ipod, iphone, PDAs etc) which are more handy and convenient to use than traditional desktop computer systems.

***Keywords:*** *Smart Phones, Graphical Passwords, Authentication, Network Security.*

## 1. Introduction

One of the major functions of any security system is the control of people in or out of protected areas, such as physical buildings, information systems, and our national borders. Computer systems and the information they store and process are valuable resources which need to be protected. Computer security systems must also consider the human factors such as ease of a use and accessibility. Current secure systems suffer because they mostly ignore the importance of human factors in security [1]. An ideal security system considers security, reliability, usability, and human factors. All current security systems have flaws which make them specific for well trained and skilled users only. A password is a secret that is shared by the verifier and the customer. "Passwords are simply secrets that are provided by the user upon request by a recipient." They are often stored on a server in an encrypted form so that a penetration of the file system does not reveal password lists [2]. Passwords are the most common means of authentication because they do not require any special hardware. Typically passwords are strings of letters and digits, i.e. they are alphanumeric. Such passwords have the disadvantage of being hard to remember [3]. Weak passwords are vulnerable to dictionary attacks and brute force attacks where as Strong passwords are harder to remember. To overcome the problems associated with password based authentication systems, the researchers have proposed the concept of graphical passwords and developed the alternative authentication mechanisms. Graphical passwords systems are the most promising alternative to conventional password based authentication systems.

Graphical passwords (GP) use pictures instead of textual passwords and are partially motivated by the fact that humans can remember pictures more easily than a string of characters [4]. The idea of graphical passwords was originally described by Greg Blonder in 1996 [62]. An important advantage of GP is that they are easier to remember than textual passwords. Human beings have the ability to remember faces of people, places they visit and things they have seen for a longer duration. Thus, graphical passwords provide a means for making more user-friendly passwords while increasing the level of security. Besides these advantages, the most common problem with graphical passwords is the shoulder surfing problem: an onlooker can steal user's graphical password by watching in the user's vicinity. Many researchers have attempted to solve this problem by providing different techniques [6]. Due to this problem, most graphical passwords schemes recommend small mobile devices (PDAs) as the ideal application environment. Another common problem with graphical passwords is that it takes longer to input graphical passwords than textual passwords [6]. The login process is slow and it may frustrate the impatient users. Graphical passwords serve the same purpose as textual passwords differing in consisting of handwritten designs (drawing), possibly in addition to text. The exploitation of smart phones like ipod and PDA's is increased due to their small size, compact deployment and low cost.

In this paper, considering the problems of text based password systems, we have proposed a new graphical password scheme which has desirable usability for small mobile device. Our proposed system is new graphical passwords based hybrid system which is a combination of recognition and recall based techniques and consists of two phases. During the first phase called Registration phase, the user has to first select his username and a textual password. Then objects are shown to the user to select from them as his graphical password. After selecting the user has to draw those selected objects on a touch sensitive screen using a stylus. During the second phase called Authentication phase, the user has to give his username and textual password and then give his graphical password by drawing it in the same way as done during the registration phase. If they are drawn correctly the user is authenticated and only then he/she can access his/her account. For practical implementation of our system we have chosen i-mate JAMin smart phone which is produced by HTC, the Palm Pilot, Apple Newton, Casio Cassiopeia E-20 and others which allow users to provide graphics input to the device. It has a display size of 240x320 pixels and an important feature of Handwriting recognition. The implementation details are out of the scope of this paper.

The structure of our paper is organized as follows. In Section II, the classification of all existing authentication methods is described. In Section III, all existing graphical password based schemes are classified into three main categories. Section IV reviews existing research and schemes which are strongly related to our work. Section V discusses the problems of all existing graphical password based schemes. In Section VI our proposed system is described in detail. In Section VII we have compared our proposed system with existing schemes by drawing out the flaws in existing schemes. Section VIII provides discussion. Finally Section IX concludes the paper.

## 2. Classification of Current Authentication Methods

Due to recent events of thefts and terrorism, authentication has become more important for an organization to provide an accurate and reliable means of authentication [14]. Currently the authentication methods can be broadly divided into three main areas. Token based (two factor), Biometric based (three factor), and Knowledge based (single factor) authentication [7], also shown in the Figure 1.

### 2.1 Token Based Authentication:

It is based on "Something You Possess". For example Smart Cards, a driver's license, credit card, a university ID card etc. It allows users to enter their username and password in order to obtain a token which allows them to fetch a specific resource - without using their username and password. Once their token has been obtained, the user can offer the token - which offers access to a specific resource for a time period - to the remote site [15]. Many token based authentication systems also use knowledge based techniques to enhance security [7].

### 2.2 Biometric Based Authentication:

Biometrics (ancient Greek: bios ="life", metron ="measure") is the study of automated methods for uniquely recognizing humans based upon one or more intrinsic physical or behavioral traits [9]. It is based on "Something You Are" [8]. It uses physiological or behavioral characteristics like fingerprint or facial scans and iris or voice recognition to identify users. A biometric scanning device takes a user's biometric data, such as an iris pattern or fingerprint scan, and converts it into digital information a computer can interpret and verify.

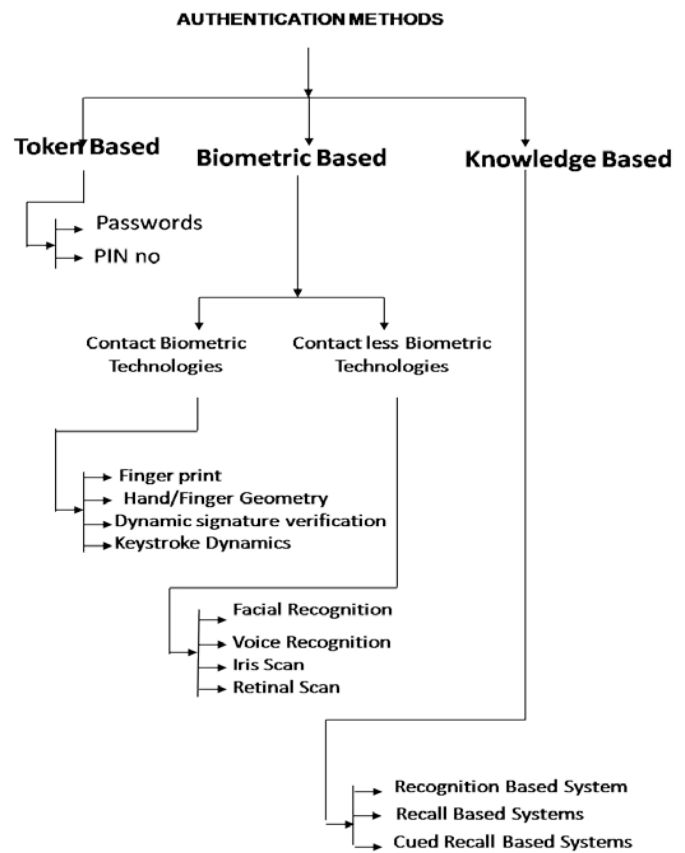

Fig. 1 Classification of Authentication Methods

A biometric-based authentication system may deploy one or more of the biometric technologies: voice recognition, fingerprints, face recognition, iris scan, infrared facial and hand vein thermo grams, retinal scan, hand and finger geometry, signature, gait, and keystroke dynamics [19]. Biometric identification depends on computer algorithms to make a yes/no decision. It enhances user service by providing quick and easy identification [20].

### 2.3 Knowledge Based Authentication:

Knowledge based techniques are the most extensively used authentication techniques and include both text based and picture based passwords [7]. Knowledge-based authentication (KBA) is based on "Something You Know" to identify you For Example a Personal Identification Number (PIN), password or pass phrase. It is an authentication scheme in which the user is asked to answer at least one "secret" question [17]. KBA is often used as a component in multifactor authentication (MFA) and for self-service password retrieval. Knowledge based authentication (KBA) offers several advantages to traditional (conventional) forms of e-authentication like passwords, PKI and biometrics [16].

## 3. Classification of Graphical Password Based Systems

Graphical based passwords schemes can be broadly classified into four main categories: First is **Recognition based Systems** which are also known as Cognometric Systems or Searchmetric Systems. Recognition based techniques involve identifying whether one has seen an image before. The user must only be able to recognize previously seen images, not generate them unaided from memory. Second is **Pure Reacll based systems** which are also known as Drwanmetric Systems. In pure recall-based methods the user has to reproduce something that he or she created or selected earlier during the registration stage. Third is **Cued Recall based** systems which are also called Iconmetric Systems. In cued recall-based methods, a user is provided with a hint so that he or she can recall his his/her password. Fourth is **Hybrid systems** which are typically the combination of two or more schemes. Like recognition and recall based or textual with graphical password schemes. Detailed classification of systems, involved in these four categories is shown in Figure 2.

## 4. Related Work

*Haichang Gao et al.* [55] have proposed and evaluated a new shoulder-surfing resistant scheme called Come from DAS and Story (CDS) which has a desirable usability for PDAs. This scheme adopts a similar drawing input method in DAS and inherits the association mnemonics in Story for sequence retrieval. It requires users to draw a curve across their password images (pass-images) orderly rather than click directly on them. The drawing method seems to be more compatible with people's writing habit, which may shorten the login time. The drawing input trick along with the complementary measures, such as erasing the drawing trace, displaying degraded images, and starting and ending with randomly designated images provide a good resistance to shoulder surfing. A user study is conducted to explore the usability of CDS in terms of accuracy, efficiency and memorability, and benchmark the usability against that of a Story scheme. The main contribution is that it overcomes a drawback of recall-based systems by erasing the drawing trace and introduces the drawing method to a variant of Story to resist shoulder-surfing.

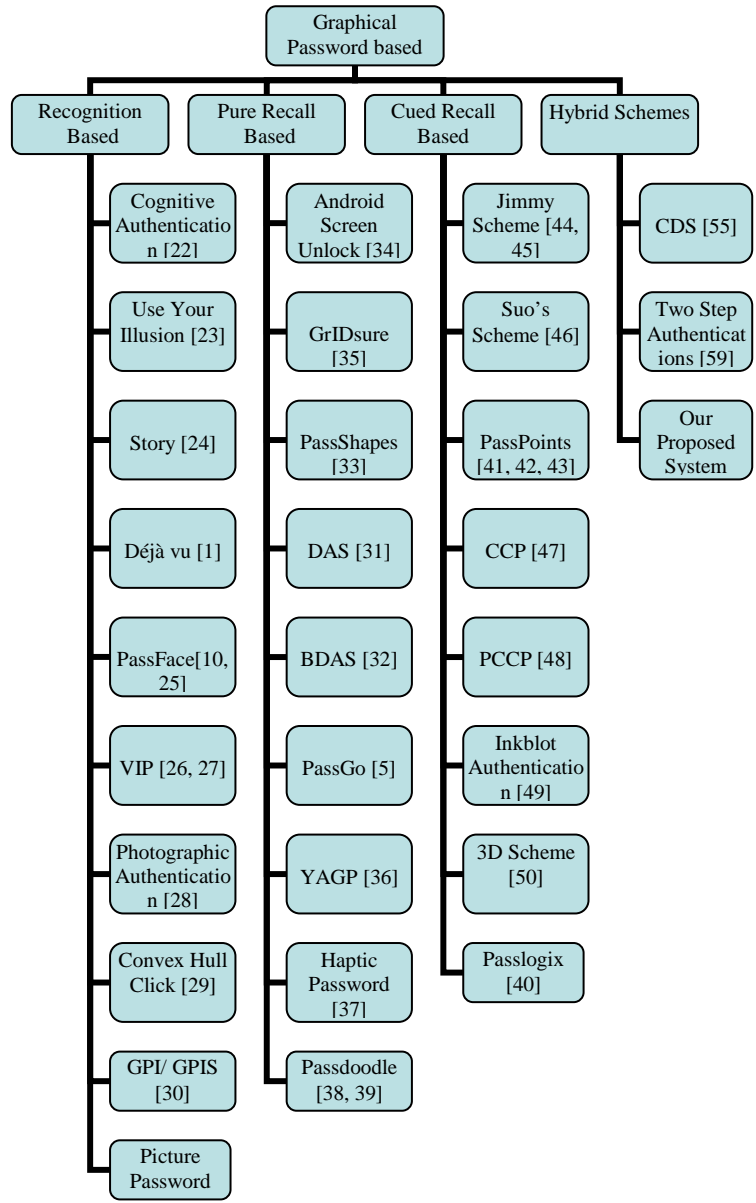

Fig. 2 Classification of Graphical Password Based Systems

*P.C.van Oorshot and Tao Wan* [59] have proposed a hybrid authentication approach called *Two-Step*. In this scheme users continue to use text passwords as a first step but then must also enter a graphical password. In step one, a user is asked for her user name and text password. After supplying this, and independent of whether or not it is correct, in step two, the user is presented with an image portfolio. The user must correctly select all images (one or more) pre-registered for this account in each round of graphical password verification. Otherwise, account access is denied despite a valid text password. Using text passwords in step one preserves the existing user sign-in experience. If the user's text password or graphical password is correct, the image portfolios presented are those as defined during password creation. Otherwise, the image portfolios (including their layout dimensions) presented in first and a next round are random but respectively a deterministic function of the user name and text password string entered, and the images selected in the previous round.

## 5. Problem Domain

There are many problems with each of the graphical based authentication methods. These are discussed below:

**5.1 Problems of Recognition Based Methods:**

Dhamijia and Perrig proposed a graphical password based scheme Déjà Vu, based on Hash Visualization technique [11]. The drawback of this scheme is that the server needs to store a large amount of pictures which may have to be transferred over the network, delaying the authentication process. Another weakness of this system is that the server needs to store the seeds of portfolio images of each user in plaintext. Also, the process of selecting a set of pictures from picture database can be tedious and time consuming for the user [7]. This scheme was not really secure because the passwords need to store in database and that is easy to see.

Sobrado and Birget developed a graphical password technique that deals with the shoulder surfing problem [3]. In their first scheme the system displays a number of pass-objects (pre-selected by user) among many other objects as shown in Fig: 3. To be authenticated, a user needs to recognize pass-objects and click inside convex hull formed by all the pass objects. They developed many schemes to solve the shoulder surfing problem but the main drawback of these schemes is that log in process can be slow.

Another recognition based technique is proposed by Man et al [63]. He proposed a shoulder-surfing resistant algorithm which is similar to that developed by Sobrado and Birget. The difference is that Man et al has introduced several variants for each pass-object and each variant is assigned a unique code. Thus during authentication the user recognize pre-selected objects with an alphanumeric code and a string for each pass-object. Although it is very hard to break this kind of password but this method still requires the user to memorize alphanumeric codes for each pass-object variants.

"Passface" is another recognition based system. It is argued by its developer that it is easy for human beings to remember human faces than any other kind of passwords. But Davis et al [12] have found that most users tend to choose faces of people from the same race. This makes the Passface password somewhat predictable. Furthermore, some faces might not be welcomed by certain users and thus the login process will be unpleasant. Another limitation of this system is that it cannot be used by those people who are face-blind [6].

**5.2 Problems of Recall Based Methods:**

The problem with the Grid based methods is that during authentication the user must draw his/her password in the same grids and in the same sequence. It is really hard to remember the exact coordinates of the grid. The problem with Passlogix is that the full password space is small. In addition a user chosen password might be easily guessable [6]. DAS scheme has some limitations like it is vulnerable to shoulder surfing attack if a user accesses the system in public environments, there is still a risk for the attackers to gain access to the device if the attackers obtained a copy of the stored secret, and, brute force attacks can be launched by trying all possible combinations of grid coordinates, ) Drawing a diagonal line and identifying a starting point from any oval shape figure using the DAS scheme itself can be a challenge for the users, and finally Difficulties might arise when the user chooses a drawing which contains strokes that pass too close to a grid-line, thus, the scheme may not be able to distinguish which cell the user is choosing.

"PassPoints" is the extended version of Blonder's idea by eliminating the predefined boundaries and allowing arbitrary images to be used. Using this scheme it takes time to think to locate the correct click region and determine precisely where to click. Another problem with these schemes is that it is difficult to input a password through a keyboard, the most common input device; if the mouse doesn't function well or a light pen is not available, the system cannot work properly [6]. Overall, with both "PassPoints" and "Passlogix", looking for small spots in a rich picture might be tiresome and unpleasant for users with weak vision.

In Viskey's scheme the main drawback is the input tolerance. Pointing to the exact spots on the picture has proven to be quite hard thus Viskey accepts all input within a certain tolerance area around it. It also allows users to set the size of this area in advance. However, some caution related to the input precision needs to be taken, since it will directly influence the security and the usability of the password. In order to practically set parameters, a four spot VisKey theoretically provides approximately 1 billion possibilities for defining a password. Unfortunately this is not large enough to prevent off-line attacks from a high-speed computer. Therefore no less than seven defined spots are required to overcome the likelihood of brute force attacks.

## 6. Proposed System

Taking into account all the problems and limitations of graphical based schemes, we have proposed a hybrid system for authentication. This hybrid system is a mixture of both recognition and recall based schemes.

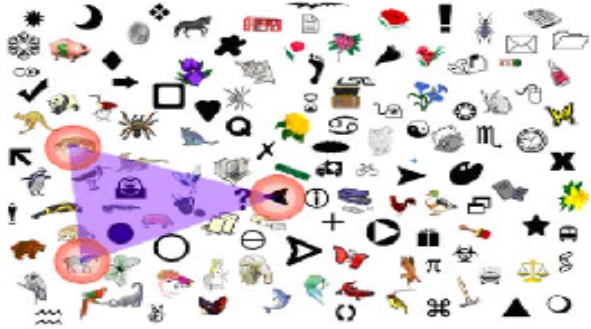

Fig. 3 A shoulder surfing resistant graphical password scheme [3].

Our proposed system is an approach towards more reliable, secure, user-friendly, and robust authentication. We have also reduced the shoulder surfing problem to some extent.

**6.1 Working of Proposed System:**

Our proposed system comprises of 9 steps out of which steps 1-3 are registration steps and steps 4-9 are the authentication steps.

**Step 1**

The first step is to type the user name and a textual password which is stored in the database. During authentication the user has to give that specific user name and textual password in order to log in.

**Step 2**

In this second step objects are displayed to the user and he/she selects minimum of three objects from the set and there is no limit for maximum number of objects. This is done by using one of the recognition based schemes. The selected objects are then drawn by the user, which are stored in the database with the specific username. Objects may be symbols, characters, auto shapes, simple daily seen objects etc. Examples are shown in Figure 4.

**Step 3**

During authentication, the user draws pre-selected objects as his password on a touch sensitive screen (or according to the environment) with a mouse or a stylus. This will be done using the pure recall based methods.

**Step 4**

In this step, the system performs pre-processing

**Step 5**

In the fifth step, the system gets the input from the user and merges the strokes in the user drawn sketch.

**Step 6**

After stroke merging, the system constructs the hierarchy.

**Step 7**

Seventh step is the sketch simplification.

**Step 8**

In the eighth step three types of features are extracted from the sketch drawn by the user.

**Step 9**

The last step is called hierarchical matching.

Graphical Representation of our proposed system is shown in Figure 5.

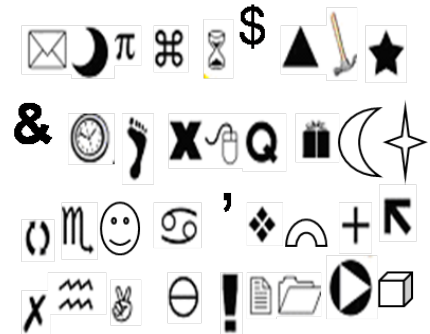

Fig. 4 Some examples of objects shown to the user

During registration, the user selects the user name and a textual password in a conventional manner and then chooses the objects as password. The minimum length for textual password is L=6. Textual password can be a mixture of digits, lowercase and uppercase letter. After this the system shows objects on the screen of a PDA to select as a graphical password. After choosing the objects, the user draws those objects on a screen with a stylus or a mouse. Objects drawn by the user are stored in the database with his/her username. In object selection, each object can be selected any number of times. Flow chart of registration phase is shown in Figure 6.

During authentication, the user has to first give his username and textual password and then draw pre-selected objects. These objects are then matched with the templates of all the objects stored in the database. Flow chart of authentication phase is shown in Figure 7. The phases during the authentication like the pre-processing, stroke merging, hierarchy construction, sketch simplification, feature extraction, and hierarchical matching are the steps proposed by Wing Ho Leung and Tsuhan Chen in their paper [13]. They propose a novel method for the retrieval of hand drawn sketches from the database, finally ranking the best matches. In the proposed system, the user will be authenticated only if the drawn sketch is fully matched with the selected object's template stored in the database. Pre-processing of hand drawn sketches is done prior to recognition and normally involves noise reduction and normalization. The noise occur in the image by user is generally due to the limited accuracy of human drawn images. [14]. A number of techniques can be used to reduce noise that includes Smoothing, filtering, wild point correction etc. Here in the proposed system Gaussian smoothing is used which eliminates noise introduced by the tablet or shaky drawing.

$$G(r) = \frac{1}{\sqrt{2\pi\sigma^2}^N} e^{-r^2/(2\sigma^2)}$$

or specifically in two dimensions

$$G(u,v) = \frac{1}{2\pi\sigma^2} e^{-(u^2+v^2)/(2\sigma^2)}$$

Where r is the blur radius (r2 = u2 + v2), and σ is the standard deviation of the Gaussian distribution.

In case, if user draws very large or a very small sketch then the system performs size normalization which adjusts the symbols or sketches to a standard size. The Stroke merging phase is use to merge the strokes which are broken at end points. If the end points are not close, then that stroke is considered as open stroke and it may be merged with another open stroke if the end point of one stroke is close to the end point of the other. The strokes are then represented in a hierarchy to simplify the image and to make it meaningful for further phases [13]. In the next step of sketch simplification, a shaded region is represented by a single hyper-stroke. After sketch simplification three types of features are extracted from the user re-drawn sketch. These features are hyper stroke features, Stroke features, and bi-stroke features.

In the last step of hierarchical matching, the similarity is evaluation the top to bottom hierarchical manner. The user is allowed to draw in an unrestricted manner. The overall process is difficult because free hand sketching is a difficult job. The order in which the user has selected the objects does matter in our proposed system i.e. during the authentication phase, the user can draw his pre-selected objects in the same order as he had selected during the registration phase. So, in this way the total combinations of each password will be $2^n - 1$, 'n' being the number of objects selected by the user as password during the registration phase.

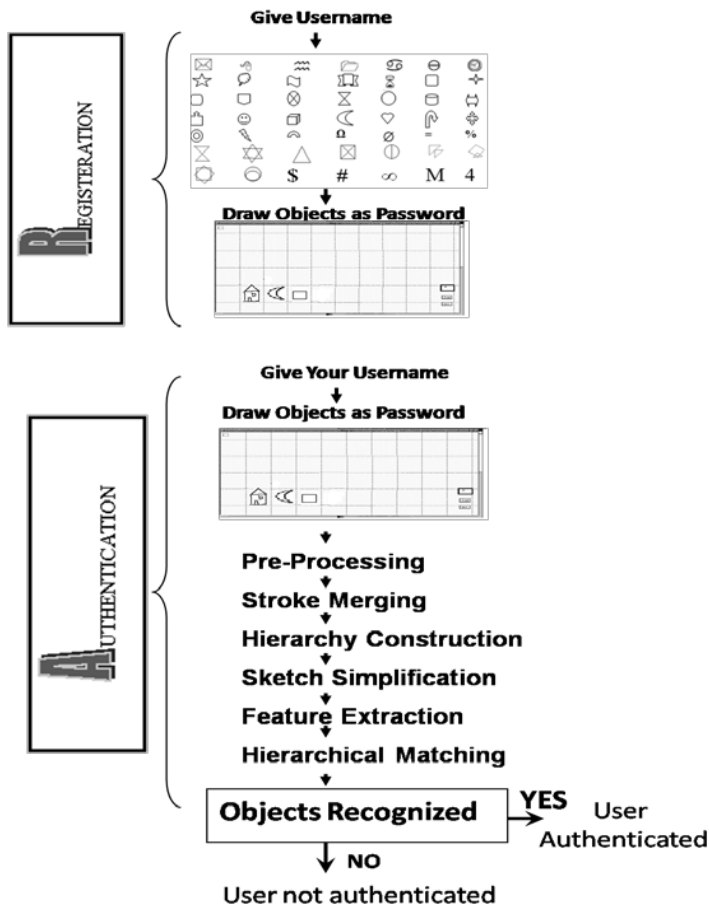

Fig. 5 Graphical Representation of Proposed System

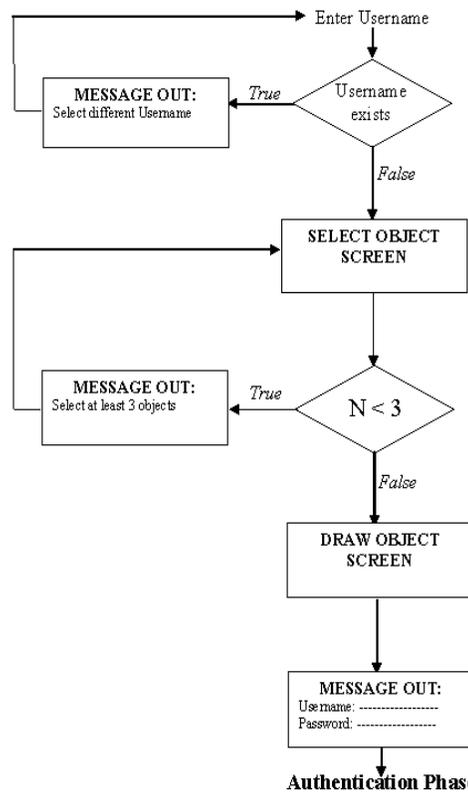

Fig. 6 Flow chart for Registration Phase

## 7. Comparison of Proposed System with Existing Schemes

Our system offers many advantages over other existing systems as discussed below:

Comparing to the "Passface" system, our system can also be used for those who are face-blind. We have used objects instead of human faces for selecting as password because later on during the authentication phase, the user has to draw his/her password and it is a much more difficult task to draw human faces than simple objects. Also we believe that as compared to human faces, objects are easier to remember which are in daily use. Our system has eliminated the problems with grid based techniques where the user has to remember the exact coordinates which is not easy for the user. Our system just compares the shapes of the objects drawn by the user during authentication.

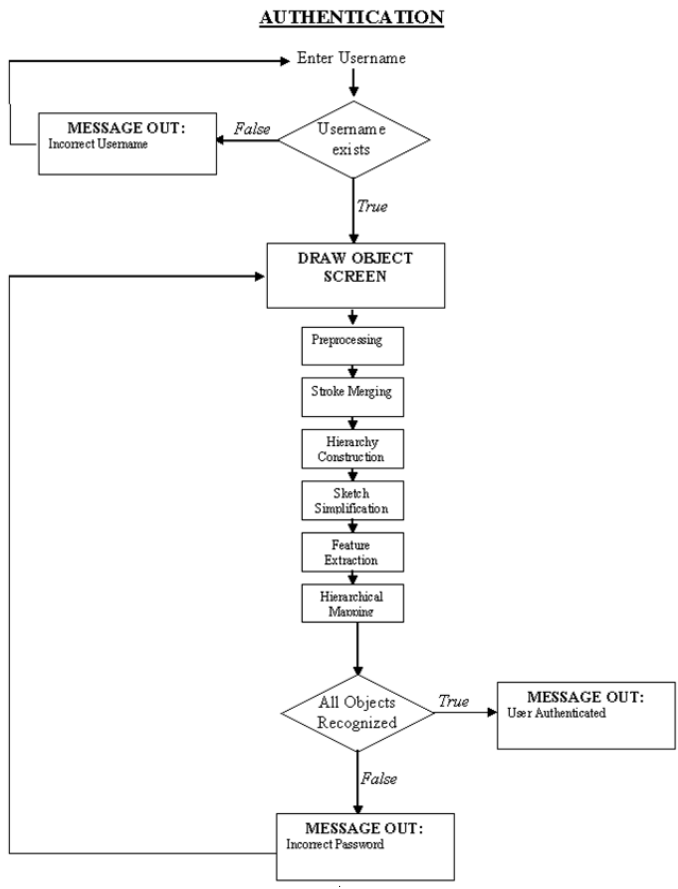

Fig. 7 Flow Chart for Authentication Phase

Our scheme is less vulnerable to Brute force attack as the password space is large. It is also less vulnerable to online and offline dictionary attacks. Since stylus is used, it provides ease to the user for drawing objects and also it will be impractical to carry out dictionary attack. Our scheme is better than Man et al scheme. This is because in his scheme the user has to remember both the objects and string and the code. In our method the user has to remember the objects he selected for password and also the way he has drawn the objects during registration.

Comparing to Van Oorschot's approach, our system is more secure since users not only select graphical password but also draw their password, making it difficult to hack. In our proposed system, even if the textual password is compromised, the graphical password cannot be stolen or compromised since the user is also drawing the graphical password. Our proposed system differs from CDS in that the user has to first select a textual password and then a graphical password, making it more secure. Comparing to Two Step Authentication system, our proposed system works in the same way as Two Step Authentication system i.e the user has to choose a textual password before choosing a graphical password but difference is that in our system during authentication, after giving the username and textual password, the user has to draw his graphical password which is matched with its stored template drawn by the user during the registration phase. This approach protects from hacking the password and prevents them from launching different attacks. Thus our system is more secure and reliable than two step authentication system. As with all graphical based systems our system will also be slow. The normalization and matching will take time. An important issue of our system is that it is somewhat user dependent during authentication. It depends upon the user's drawing ability. Thus, the system may not be able to verify the objects drawn by the user and as a result the actual user may not be authenticated.

The possible attacks on graphical passwords are Brute force attack, Dictionary attacks, Guessing, Spy-ware, Shoulder surfing and social engineering. Graphical based passwords are less vulnerable to all these possible attacks than text based passwords and they believe that it is more difficult to break graphical passwords using these traditional attack methods. Our System is resistant to almost all the possible attacks on graphical passwords. The comparison of our system to existing schemes and systems in resisting attacks on graphical passwords is shown in table 1.

## 8. Conclusion & Future Work

The core element of computational trust is identity. Currently many authentication methods and techniques are available but each with its own advantages and shortcomings. There is a growing interest in using pictures as passwords rather than text passwords but very little research has been done on graphical based passwords so far. In view of the above, we have proposed authentication system which is based on graphical password schemes. Although our system aims to reduce the problems with existing graphical based password schemes but it has also some limitations and issues like all the other graphical based password

Table Comparison Of Graphical password Schemes resistant to different Attacks

| Graphical Password Schemes/ Systems | Type of Scheme | Resistant to Possible Attacks | | | | | |
|---|---|---|---|---|---|---|---|
| | | Brute Force Attack | Dictionary Attack | Guessing Attack | Spy-ware or Naïve Key logging | Shoulder Surfing Attack | Phishing Attack or Social Engineering |
| Blonder's Scheme [62] | Recognition Based | Y | N | Y | N | Y | N |
| DAS [31] | Pure Recall Based | Y | N | Y | N | Y | N |
| BDAS [32] | Pure Recall Based | N | - | - | - | - | - |
| Qualitative DAS [65] | Pure recall Based | N | - | - | - | - | - |
| Syukri Algorithm [64] | Pure recall Based | N | Y | Y | N | Y | N |
| PassPoints [41, 42, 43] | Cued Recall Based | Y | N | Y | N | Y | N |
| PassFace [10, 25] | Recognition Based | Y | Y | Y | N | Y | N |
| PassGo [5] | Pure Recall Based | Y | - | - | - | - | - |
| Passlogix [40] | Cued Recall Based | Y | N | Y | N | Y | N |
| PassMap [66] | Pure Recall Based | Y | N | - | N | Y | N |
| Passdoodle [38, 39] | Pure Recall Based | N | - | - | - | - | - |
| Viskey SFR | Pure Recall Based | Y | N | Y | N | Y | N |
| Perrig and Song [11] | Recognition Based | Y | N | Y | N | Y | N |
| Sobrado and Birget [3] | Recognition Based | Y | N | Y | N | N | N |
| Man et al Scheme [63] | Recognition Based | Y | N | N | Y | Y | N |
| Picture Password Scheme [60] | Recognition Based | Y | N | Y | N | Y | N |
| CDS [55] | Hybrid | - | - | - | - | Y | - |
| WIW [57] | Recognition Based | - | - | - | - | Y | - |
| Association based scheme [58] | Recognition Based | - | - | - | - | Y | - |
| Déjà Vu [1] | Recognition Based | Y | - | Y | - | - | - |
| Haptic Password Scheme [37] | Pure Recall Based | - | - | - | - | Y | - |
| YAGP [36] | Pure Recall Based | Y | - | Y | - | Y | - |
| Photographic Authentication [28] | Recognition Based | - | Y | - | - | - | - |
| Two Step Authentication [59] | Hybrid | - | - | - | Y | N | Y |
| Our Proposed System | Hybrid | Y | Y | Y | Y | Y | Y |

Note: Y= Yes resistant to attack   N=No not resistant to attack

techniques. To conclude, we need our authentication systems to be more secure, reliable and robust as there is always a place for improvement. Currently we are working on the System Implementation and Evaluation. In future some other important things regarding the performance of our system will be investigated like User Adoptability and Usability and Security of our system.

**Acknowledgment**

The authors wish to acknowledge the anonymous reviewers for valuable comments.

**Wazir Zada Khan** is currently with School of Computer Science, Jazan University, Kingdom of Saudi Arabia. He received his MS in Computer Science from Comsats Institute of Information Technology, Pakistan. His research interests include network and system security, sensor networks, wireless and ad hoc networks. His subjects of interest include Sensor Networks, Wireless Networks, Network Security and Digital Image Processing, Computer Vision.

**Dr. Muhammad Y Aalsalem** is currently dean of e-learning and assistant professor at School of Computer Science, Jazan University. Kingdom of Saudi Arabia. He received his PhD in Computer Science from Sydney University. His research interests include real time communication, network security, distributed systems, and wireless systems. In particular, he is currently leading in a research group developing flood warning system using real time sensors. He is Program Committee of the International Conference on Computer Applications in Industry and Engineering, CAINE2011. He is regular reviewer for many international journals such as King Saud University Journal (CCIS-KSU Journal).

**Dr. Yang Xiang** is currently with School of Information Technology, Deakin University. He received his PhD in Computer Science from Deakin University. His research interests include network and system security, distributed systems, and wireless systems. In particular, he is currently leading in a research group developing active defense systems against large-scale distributed network attacks and new Internet security countermeasures. His recent research has been supported by the Australian Research Council (ARC), the University, and industry partners. Dr. Xiang has published more than 100 research papers in international journals and conferences. He has served as Program/General Chair for many international conferences such as ICA3PP 11, IEEE/IFIP EUC 11, TrustCom 11, IEEE HPCC 10/09, IEEE ICPADS 08, NSS 11/10/09/08/07. He has been PC member for many international conferences in distributed systems, networking, and security. He is regular reviewer for many international journals such as IEEE Transactions on Parallel and Distributed Systems, IEEE Transactions on Dependable and Secure Computing, IEEE Transactions on Information Security and Forensics, IEEE Communications Letters, and IEEE Journal on Selected Areas in Communications. He is on the editorial board of Journal of Network and Computer Applications.